# Deployment of Stream Control Transmission Protocol (SCTP) to Maintain the Applications of Data Centers

Fatma Almajadub, Eman Abdelfattah, Member IEEE, and Abdul Razaque, Senior *Member, IACSIT*

*Abstract*— with developments of real-time applications into data centers, the need for alternatives of the standard TCP protocol has been prime demand in several applications of data centers. The several alternatives of TCP protocol has been proposed but SCTP has edge due to its several well-built characteristics that make it capable to work efficiently. In this paper, we examine the features of SCTP into data centers like Multi-streaming and Multi-Homing over the features of TCP protocol.

In this paper, our objective is to introduce internal problems of data centers. Robust transport protocol reduces the problems with some extend. Focusing the problems of data centers, we also examine weakness of highly deployed standard TCP, and evaluate the performance of SCTP in context of faster communication for data centers. We also discover some weaknesses and shortcomings of SCTP into data centers and try to propose some ways to avoid them by maintaining SCTP native features. To validate strength and weakness of TCP and SCTP, we use ns2 for simulation in context of data center. On basis of findings, we highlight major strength of SCTP. At the end, we Implement finer grain TCP locking mechanisms for larger messages.

## I. INTRODUCTION

The data centers represent the foundation of the Internet and computer services specially E-business service and computing with the high performance. Nowadays, web service development is based on the increased size and the complexity of the processing data. It is clear that the data centers continue to grow with performance requirements, availability requirement and developed requirements. Hence this remarkable growth in the data centers, have motivated number of researchers to improve data transfer. Most of work is done on front sides [1]. Due to the heavy load of network traffic; TCP/IP/Ethernet fails to control the congestion in data centers over the network. Thus it causes of massive loss of confidential data and wastage of sources. Although TCP/IP/Ethernet are completely deployed into the data centers and work as stacks, they do not have capacity to control the huge amount of data. For example, IBA is designed to work and act as a universal data center, but it is getting acceptance in only certain areas. Another example, Fiber channel, which is designed for the specialized networks like Infiniband (IBA), is spread for high-end system of (IPC) inter-process communication. On the other hand, the technology of Ethernet continues to remain the best choices to education, e-business, big markets etc. The reason for that is related to many factors such as the incompatibility factor at the level of connector between IBA and Ethernet or familiarity factor [2].

In addition, there should be protocol to carry all kinds of traffic into data centers even though the storage development of IP protocol like Internet Small Computer System Interface (ISCSI), TCP/IP/Ethernet. It should also have capability to transfer 10 GB/S and handle the problem of Homework (HW) protocol. However, all previous studies until now refer and expect that IP protocol is well scaled into the data center, but there are several fuzzy things and questions about transparency of TCP protocol that is connected with IP protocol for supporting the applications of data centers. For example, the demanding of high data rate, low latency, high robustness, high availability and so on. Since the ambiguity and the weakness of TCP protocol are well known, it is impossible to create or do considerable changes on TCP protocol [3] & [4].

However, there are several alternative variants of TCP which are used in the areas where TCP cannot work. Fiber Channel Protocol (FCP) which is preferred to use on Storage area networks (SANs) and also work with real-time applications but TCP is unsuitable for such type of applications. Another example, SCTP which is a connection-oriented transport protocol and another IP protocol that provides reliable stream oriented services similar to TCP. SCTP is especially designed to be used in situations where reliability and near-real-time considerations are important as well as it is designed to run over existing IP/Ethernet infrastructure. [5].

Moreover, SCTP was designed for support of Signaling System 7 (SS7) layers like (Message Transfer Part) MTP2 and MTP3. It also works with SS7 and voice channel over internet protocol (VoIP) network. Therefore, SCTP protocol is the best for data center. [5]. SCTP has many promising features including the flexibility, robustness, and extensibility [6], [7]. Therefore, we introduce the study of SCTP congestion mechanism into data centers and the impact of some optimizations that we have studied to develop SCTP and reaching to the way that maintain the applications into data Centers. Furthermore, we demonstrate all the sides of protocol.

The paper is organized as follows: In section 2: we present the features of SCTP for data center requirements. In section 3: the evaluation of data centers and WAN Environments are discussed. In section 4: The features of TCP and SCTP are examined. In section 5: performance enhancement of SCTP is highlighted. In section 6: simulation results and finally section 7, concludes the paper and future work.

## II. FEATURES OF SCTP FOR DATA CENTER REQUIREMENTS

Although TCP protocol has many features, it was not designed to use for the data centers. Also, some of its weaknesses become acute and need to study in some environments as we discuss in this paper. For that there was the need of protocol SCTP. From other side, SCTP adopts congestion window/flow control scheme of TCP except for some minor differences [8], [9], this makes SCTP identical from TCP protocol in the behaviors of its congestion and flow control.

On the other hand, SCTP has provided many improvements over TCP as the following:

Multi-streaming: SCTP connection can have multiple streams; each of them specifies a logical channel. Although the flow and congestion control are still on the basis of each connection, the streams can be exploited for many purposes like giving the higher priority to messages and more [10], [11].

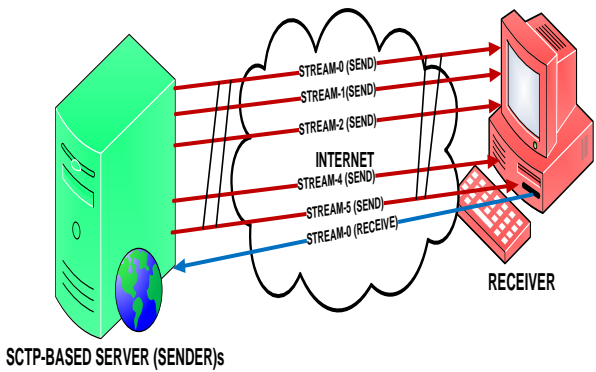

Figure 1: Multi-streaming process of SCTP

Multi-homing: SCTP connection can define multiple "endpoints" on each end of the connection that increases level of connection to handle with errors. If primary connection fails then, the sender selects alternate primary connection for forwarding data until it is restored shown in figure 2.

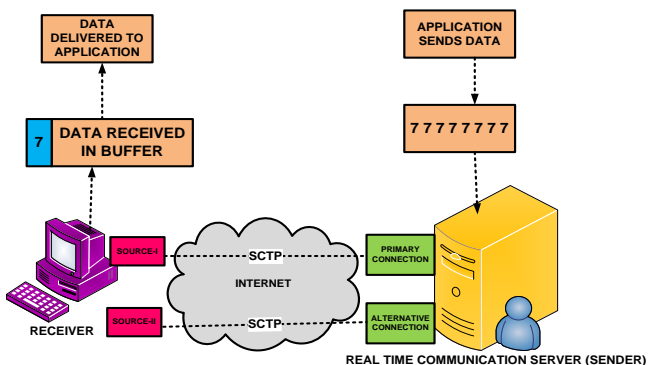

Figure 2: Multi-homing process of SCTP

One of the promising features of SCTP is to handle the denial of service attack. It sets up SCTP connection including 4 messages (4-way handshaking) and avoiding propagation of any message at the endpoint until it has ensured that the other end is interested in setting up connection [12] given in figure 3.

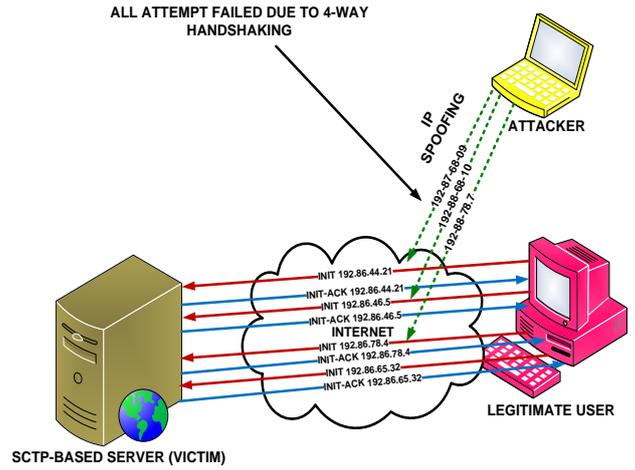

Figure 3: 4-way handshaking process of SCTP

Flexibility in-order delivery of packets causes the reduction of latency. Thus, each SCTP stream provides well organized in-order delivery [13], [14].

Robust connection: SCTP connection maintains a verification tag that is provided for each subsequent data transfer so that it is robust against tapping and errors. This is vital within data center for transferring high data rates [15], [16].

## III. EVALUATION OF DATA CENTERS AND WIDE AREA NETWORK (WAN) ENVIRONMENTS

When we compare between data centers and WAN environments, we find many differences. However, we focus on internal side of data center and how to multiply clusters of connection. Some of these differences as following:

1. Data centers have completely different requirements from the requirements of general WAN.
2. The flow of data centers adapts automatically with the environment and provides the highest throughput in highly congested network. Also the flow is fair with other competing flows.
3. Data centers require higher levels of robustness, availability, flexibility in ordering.
4. Data centers as compare to WAN, has the characteristics of communication that includes less variable round-trip times (RTTs), higher data rates, higher installing capacity, less congestion and very low latency requirement.
5. Data centers have architectural protocol that is altogether different from architectural protocol of WAN. By examining the protocols of data centers such as Myrinet or IBA, we observe the improved throughput is less important than overhead of a protocol processing. In addition, the communication latency must be a low and the most significant for protocol architecture.
6. Data centers work with CPU utilization for a given throughput, but WAN environments don't give interest to CPU utilization.

7. Data centers demand the higher levels of robustness and availability. So that requirements of robustness increase with certain speeds.

From the previous comparisons we can incur the following conclusions:

(a) Preferably, the implementation of 0-copy, which we send and receive that is accessible for this purpose based on standard.
(b) Copies of memory-to-memory (M2M) are obtained for sending large data in the cost of processor bus BW, CPU cycle, latency and memory controller BW. In addition, remote direct memory access (RDMA) is getting wide acceptance as proficient 0-copy transfer protocol [17], [18]. However, an efficient deployment of RDMA is complex on byte stream abstraction.
(c) Implementation of protocols mostly relies on multiple copies of (M2M) that is considered as tool for suitable interfacing of various software layers.

Therefore, SCTP can be interfaced and also compatible with RDMA. As result of this evaluations, there are other differences between WAN and data center environments. We should address them in terms of optimizing SCTP for data centers and its use shown in figure 4.

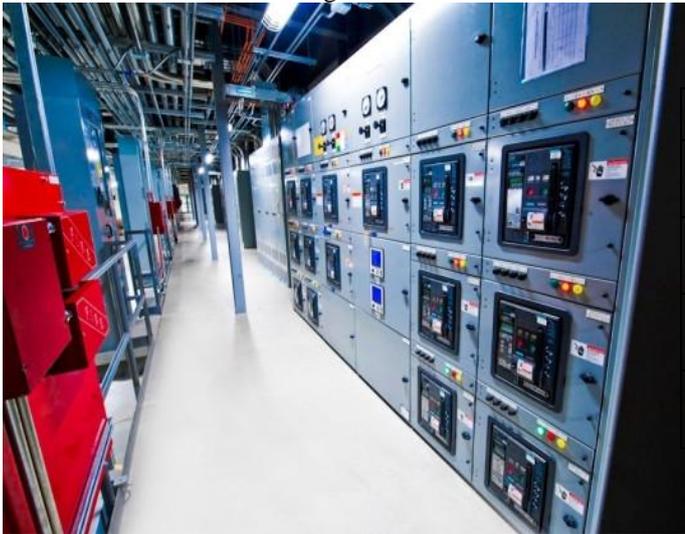

Figure 4: Data center for processing data communication

## IV. EXAMING THE FEATURES OF TCP and SCTP

There is big difference between TCP and SCTP protocol. SCTP protocol comes with extra promising features and it is considered as fine open-source implementation. The used open-source is supported with Linux 2.6.16 Kernel. We have chosen and studied this mode for experiments due to non-availability of equipments for free System Demon (BSD) like Emon, SAR, Oprofile, as well as problem of running and familiarity with Linux. Furthermore, we conduct different tests that comprise of unidirectional data transfer such as bit like test transmission control protocol (TTCP) that has edge over tools like file transfer protocol (FTP).

We also deploy version of IPerf tool that comes with Linux Kernel (LK-SCTP) allocation. Since, IPerf doesn't have multi-streaming capacity but it sends messages of given size such as "back to back". However, multi-streaming tests are conducted using small traffic generators [19]. We observe that LK-SCTP is required to work on two machines running on R.H 9.0 with 2.6 Kernel [20]. These machines have 512 KB second level cache and processor with speed of 2.8 GHz Pentium-IV including Intel GB NICs. The message of TCP doesn't require additional work such as known of the both ends of message. But that is required with SCTP protocol for recognition the boundary of message.

In addition, SCTP protocol works with a cyclic redundancy check-32 (CRC-32), where checksum calculation is CPU intensive. Although CRC-32 increases the protocol processing cost by 24% on the sender side and 42% on the receiver side. CRC-32 achieves the high speed .Therefore, one of important differences between TCP and SCTP is HW offloading. Whereas TCP protocol is provided with NICs to have the capability of TCP transport segmentation offload (TSO) and checksum offload, SCTP protocol does not have that features. For that we do not use the cyclic redundancy code (CRC) for SCTP implementation. STCP algorithm works as follows:

### A. Base Performance Comparisons:

The following table appears some of comparisons between TCP and SCTP protocol:

TABLE 1: 12 KB TRANSFERS, 1 CPU, 1 CONNECTION

| Parameters | Total CPI | Path-length | 2ndL MPI | CPU unit | T-put (Mb/s) |
|---|---|---|---|---|---|
| TCP Send without TSO & Checksum | 6.45 | 24910 | 0.04275 | 62.4 | 1394 |
| SCTP Send without TSO & Checksum | 4.41 | 91059 | 0.0264 | 143.2 | 1375 |
| TCP Receive without TSO & Checksum | 5.835 | 30885 | 0.08145 | 60.5 | 1376 |
| SCTP without TSO & Checksum) | 5.88 | 53880 | 0.0501 | 105.1 | 1356 |

The comparison in the table1 is on the basis of a single connection running over the GB NIC and pushing 12 KB packets as fast as possible under zero packet drops. Therefore, SCTP is configured with only one stream with 12 KB as the size of the receive windows. Also, we found SCTP protocol can result the same throughputs as TCP.

The performance includes the following major parameters:

*1. CPU utilization*

SCTP-send is 3.7X processing intensive as compare with TCP; its send is in terms of CPU.

- *Average CPU cycles per instruction (CPI)*

The CPI numbers focus on the nature side of the inefficiency. We found the overall CPI is only 68% Since SCTP works on executing 3.7X instruction and that is simpler and has better caching behavior. That is like the instructions of TCP protocol.

- *Path-length or number of instructions per transfer (PL)*

The PL numbers focus on the nature side of the inefficiency.

- *No of cache misses per instruction in the highest level cache (MPI).*

The MPI focuses on the nature side of the inefficiency.

AS result, the performance includes the previous major parameters that make SCTP is more efficient about 3.7X than TCP and that is on the receiver end. In addition, the results show SCTP needs less work load based on the basic of TCP.

We also found that the measure of performance efficiency is the throughput rather than the CPU utilization. Although we have presented the data transfer with the large sizes (12 KB). The operations impact of performance (M2M copy) is to obtain the performance of the applications such as ISCSI that is shown in the table 1. It is also important to note performance with the small sizes of data transfer, such as, 128 byte or less where the processing of packets Confuse the CPU for the TCP and SCTP protocol which is given in Table 2.

TABLE 2: 128 B TRANSFERS, 1 CPU, 1 CONNECTION

| Case | T-put 128 KB | T-put 256 KB |
|---|---|---|
| TCP Send without TSO & Checksum | 132 | 264 |
| SCTP Send without TSO & Checksum) | 102 | 204 |
| TCP Receive without TSO & Checksum | 262 | 524 |
| SCTP Receive without TSO & Checksum | 219 | 438 |

*B. The Default Setting of TCP and SCTP*

We observe that there are many differences between standard TCP and SCTP protocol, but all of them are based on the size of window and collection of data. Therefore, we determine by default that congestion window allows sending maximum transmission unit MTU. Thus SCTP protocol does not wait for more arrival packets, but it builds the packet from the application messages which are available. In addition, we make SCTP to provide a NO-DELAY option, when we make it by default. Furthermore, SCTP is the message oriented and provides the capability to bundle the chunks [21].

We also observe the behavior of TCP that is considered by default as a byte-stream oriented protocol. TCP accumulates only one data of MTU values. It calls IP datagram before sending the packets. Therefore, the undesirable delay may be counted, if data is not arrived as a continuous stream from the application layer. Therefore, we make TCP to provide a NO-DELAY option which by default it is turned off.

*1. The expected results of TCP and SCTP on default setting*

As result, TCP outperforms SCTP because of fewer data structure manipulations. As the following:

a. TCP is found more efficient than SCTP.
b. TCP appears to perform better than SCTP.
c. In the data center, SCTP performs better than TCP because SCTP has capability of handling more data than TCP. Data center deals with large amount of data on daily basis.
d. SCTP with the chunk bundling must be enabled because it only works within the available data.
e. According to the previous points, the performance of SCTP is worse than TCP and this was assumed in the Table 2 in the second column where the windows size is 128KB and 256KB.

*C. Multi-Streaming Feature of SCTP VS TCP:*

Table 3 shows comparisons between SCTP and TCP protocol on basis of 4 connections and association with 2 streams. On the same NIC.

TABLE 3: 2.56 KB TRANSFERS WITH 2 CPUs, ALL CONNECTIONS

| Parameter | Total CPI | path length | 2ndL MPI | CPU utilization | Th-put Mb/sec |
|---|---|---|---|---|---|
| TCP Send with 4 connection | 10.68 | 8675 | 0.0769 | 79.2 | 1705 |
| SCTP Send 4 association with 2 stream | 10.2 | 23504 | 0.0876 | 198 | 1776 |
| TCP Receive with 4 connection | 8.92 | 7890 | 0.1204 | 69 | 1794 |
| SCTP 4 association with 2 stream | 12.4 | 15604 | 0.01024 | 129 | 1780 |

*1. The default setting of TCP and SCTP:*

a) We look over the scenario of a single NIC with 4 associations or 4 connections based on that the streams of a single association or connection cannot be split over multiple NICs.
b) We have changed the transmission size from the 12 KB down to 2.56 KB for avoiding the single NIC. In addition, we also used a configuration of a dual processor DP for making sour that the CPU does not become the problem of the bottleneck.
c) We make it by default using same CPU utilization for multi-streaming that is better than multi-association.
d) We provide multi-streaming in SCTP which is lightweight nature and different from the associations or connections. Based on that the flow and congestion control in SCTP which is available for all the streams, are the more easily implementation.

*2. The expected results of TCP and SCTP on the default setting:*

The following results are almost correct for both the sending and receiving. In addition, these results are based on that the streams are the same weight like associations or connections as well as the streams aren't able to make the CPU arrives to 100% of utilization. The results as the following:

a) The arriving rate of SCTP is higher. So that the chunks of SCTP must be removed where simultaneous processing of 2 streams initiate. Otherwise it causes serious problem, and that is considered as fundamental shortcoming for the feature of stream.
b) The structure of transmission control block (TCB) must be changed along with finer granularity locking for relieving the problem which is caused by the resulting lock contention. It limits severely the stream throughput. That problem of the implementation is at the function of Sending of LK-SCTP which opens the socket at time. Message is also received by the IP-layer and locks the socket at the beginning of the function. According to given streams shortcomings are created in the side of protocol specification and in the side of implementation.

c) SCTP is less efficient in the single connection case though the SCTP as TCP are able to execute and achieve almost the same throughput.
d) According to pervious point the structure of TCP and the handling for SCTP have some deficiencies which were explained on the experiments as well.
e) Overall throughput of SCTP with two streams over 4 associations or 4 connections is about 52% and is less than that for two associations.
f) The CPU utilization of SCTP with two streams is also about 52% and lowers than for the 4 associations or 4 connections.

## V. PERFORMANCE EVALUATION ENHANCEMENT OF SCTP

We showed the performance improvements from enhancements as well as we compared the parts of SCTP with TCP. These parts which make these enhancements are difficult. Therefore, we showed the implementation of LK-SCTP based on the viewpoint of efficiency and specify some parts for performance enhancements.

### A. LK-SCTP approach:

Figure 5 shows approach of LK-SCTP to chunking and chunk bundling. And it is as the following:
1. The message, which is specific for each user message, contains the list of chunks. That depends on that approach of LK-SCTP, which maintains 3 data structures to manage the chunk.
2. The first structure is free only when all chunks which belong to it, are acknowledged by the remote endpoint.
3. The two other data structures, which are specified and freed by LK-SCTP, manage each chunk as the following
    a. The first structure contains the actual chunk buffers and the chunk header.
    b. The second structure contains pointers to chunk buffers and some different data.
4. Many small data structures are maintained by the implementation. They are executed by specified and de-specified of memory. The chunk is copied to the final buffer after it is processed by many procedures and routines based on that LK-SCTP approach so that before it copies variables and values to the final destination; it initializes the local variables with values.

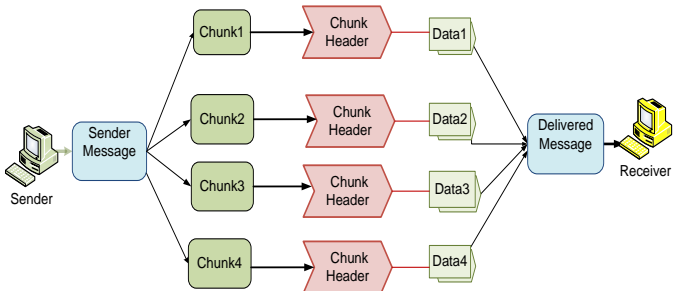

Figure 5: Process of sending message

### B. The results of LK-SCTP approach:

Before transmitting the data on the wire, it is resulted three copies of M2M which include direct memory access (DMA) of data into the NIC buffers for the sending in the wire. As well as passing control to NIC by bundling chunks into a MTU packet and retrieving the data from the user buffer as well as investing in the data message structure. The LK-SCTP implementation can be speeded up by the using of the Cut Down technique on the M2M copies for large messages as well as the using of avoiding technique to dynamic memory allocation/de-allocation in ring buffers and the using of avoiding technique to chunk bundling only as it is appropriate.

The ideal implementation uses pre-allocation and helps to reduces number of copies. Therefore, while we fragment the large user data, we decide whether a chunk of a given data can be bundled together with other chunks or not. If not, we designate this chunk as a full chunk and prepare a packet with one chunk only. In addition, we have worked to eliminate 1 copy for messages that are larger than 1024 bytes by bundling chunks into a MTU packet and retrieving the data from the user buffer. Also, we did that for the smaller message by turning out the default 2-copy path to be shorter. AS result, the current beginning of 1024 bytes was achieved and may shift as extra optimizations.

We found many small control packets or SACK packets by the using of the ethereal tool to look at the packet sequentially and LK-SCTP which works on processing the large amount of these packets on the sender ends and receiver ends. Also, we found 2 SACKs packets are sent by LK-SCTP rather than one SACK and this is equal to one SACK per packet. The first SACK packet is sent when the packet is received to the application and the second SACK packet is sent when the packet is delivered to the application.

We determine the SACK processing overhead of SCTP is more expensive due to multi-streaming features, chunking, and immature implementation. Therefore, the frequency of SACKs in SCTP is higher because SCTP lacks acknowledgment. For that we make the frequency of SACKs packets to per 7 packets and insure sending it either on data delivery or on data receives when delivery is not possible because of missing packets.

On the side of the size and layout of connection descriptors which is called TCB; we found that the size of the connection or association structure was the bigger size at 10 KB for SCTP Whereas TCB size is equal to 1024 bytes for TCP. In addition, we found large TCB sizes aren't desirable for caching efficiency and processing complexity.

The maximum burst size MBS which is the final feature of SCTP was considered for optimization. That controls on the maximum number of data chunks, which sent on any given stream before waiting for an acknowledgement.

## VI. SIMULATION RESULTS

We show comparison between performances of SCTP with w/o optimizations against TCP. We have estimated a well optimized implementation and certain protocol changes that should be close the performance of TCP.

### A. Performance Impact of Optimizations over TCP and SCTP:

SCTP should be able to provide best performance than TCP. Figure 6 compares between SCTP CPU utilization and TCP with and without optimization with 12KB data transmission.

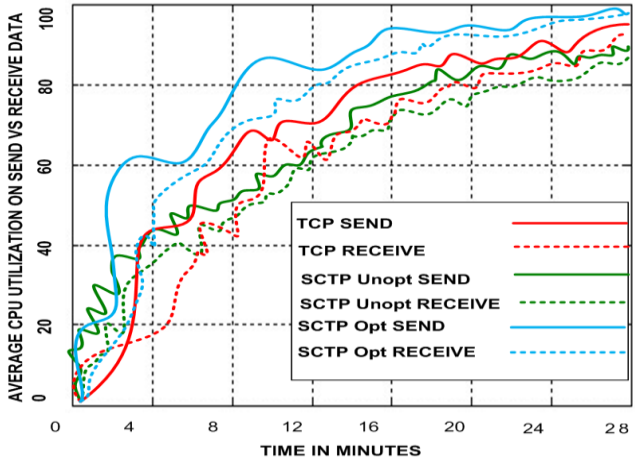

Figure 6 Average CPU utilization for 12 KB transfers

We have obtained desired throughput that is about the same (~1705Mb/sec). Furthermore, for SCTP; the optimizations drop CPU utilization from1.16x to 3.7x and SCTP receive utilization also improves from 1.9x down to about 1.42x.

Figure 7 shows the scaling of SCTP as a number function of connections. Also, we notice each new connection which is carried over a separate GB NIC for ensure that the throughput is not limited by the NIC. It considers that the original SCTP scales with number of connections and the optimizations bring it closer to TCP scaling.

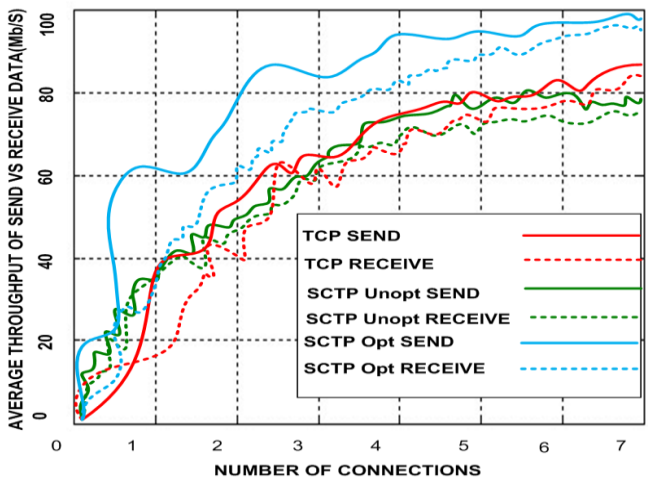

Figure 7 Throughput scaling with multiple connections

The CPU With three simultaneous connections becomes the problem of a bottleneck for both TCP and SCTP; thus, the scaling from 2 to 4 connections is bad and poor for TCP and SCTP protocols.

Figure 8 shows the SCTP throughput for small packets which are around 128KB. The performance with these packets depends on the receiving window size and the NO-DELAY option. We found that the results are shown in this figure are for NO-DELAY on and receiver window size of 128 KB to 256 KB. Therefore, SCTP and TCP throughputs were already comparable and the throughput improves, but not that much. Optimized SCTP sending throughput is actually higher than that for TCP. We also found when the size of a receiver window 128 KB, TCP continues to outperform optimized SCTP.

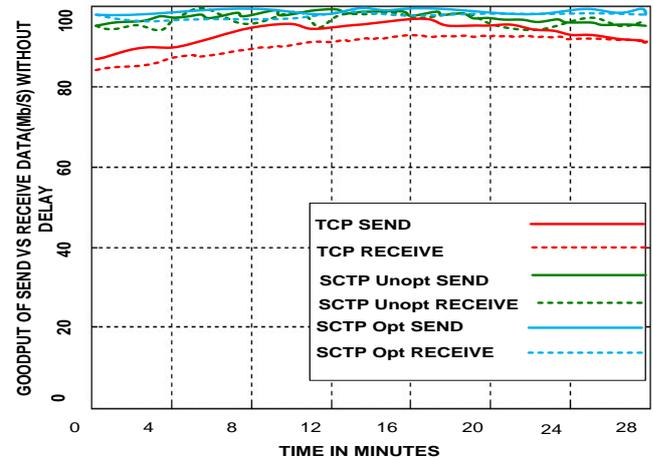

Figure 8 Throughput comparisons with 128 Bytes packets

### B. Evaluation of SACK and ACK in Data Centers over TCP and SCTP

The comparisons are depending on the attraction of SACKs which is reported of individual gaps whereas the missing packets are retransmitted.

1. The SACK structure is designed for arbitrary lists of gaps and only leads to overhead. If SACKs are sent for every two packets, it will report at most two gaps, and usually no more than one gap. Also, within data center environment, a reduced SACK frequency is an obvious optimization for data centers and the gap reporting is less efficient since a single gap will appear very rarely.
2. More significant point is that SACKs aren't desirable in a data center.

    a. We observe there is no need to use any buffers to keep unacknowledged data on receiving side without SACKs. It can also be very cost savings at high data rates.

    b. Round-trip times (RTTs) within data center is small and extra retransmissions is done if SACKs are not used, that is the more beneficial.

    We try to make some changes in SACK mechanism of SCTP protocol to allow it to emulate go-back-N (GBN) type of SACK protocol as well. We base on the further study to the relevance of SACK in the data center environment. In addition, we did this implementation for the experimentation and expected that these implementations can be done simply and efficiently. But it requires significant changes to the protocol SCTP.

Figure 9 shows and illustrates the following: TCP and SCTP performance under random packet losses.

3. The achieved throughput for a GB NIC for 12 KB data transfers. Because several differences in the congestion control algorithm which used by the SCTP and TCP protocols, SCTP performs better than TCP under low drop rates and worse for high drop rates. In addition, we expected that due to a reduction in SACK frequency which is detrimental to throughput performance at high drop rates and it is desirable at lower drop rates.

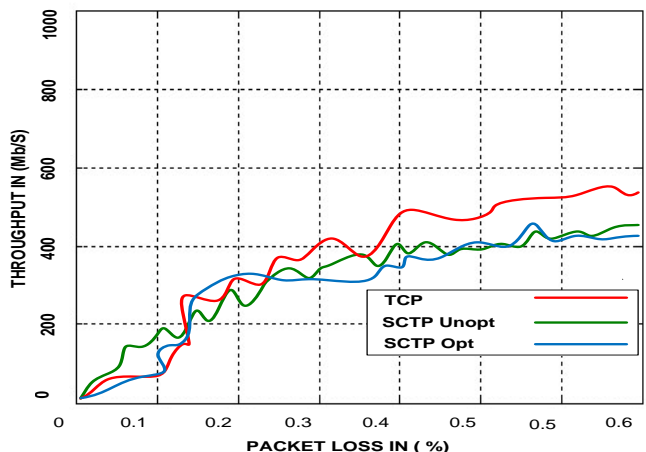
Figure 9 Throughput comparisons with the Packet loss

4. Maximum throughput is with 12 KB of message size for SCTP. It is also set with both options: the SACK and emulated GBN options. Therefore, we observe that the CPU utilizations for SACK and GBN options are not reported and are almost identical across the board; thus, the comparison of a direct throughput is correct and describes and reflects the differences between the two cases.

Due to that, the difference is clearly for a function of round-trip times (RTT). When, the RTT values are consistent with the data center environment by around 102 microseconds within this value of the product of the bandwidth-delay at 2 GB/sec is only 12KB, or less than one user message. Also, the extra retransmission overhead is more than compensated by simpler processing with ideal and nominal SACK rate per 12KB message and moderate drop rates. Depending on that, we observe as it is expected as follows:

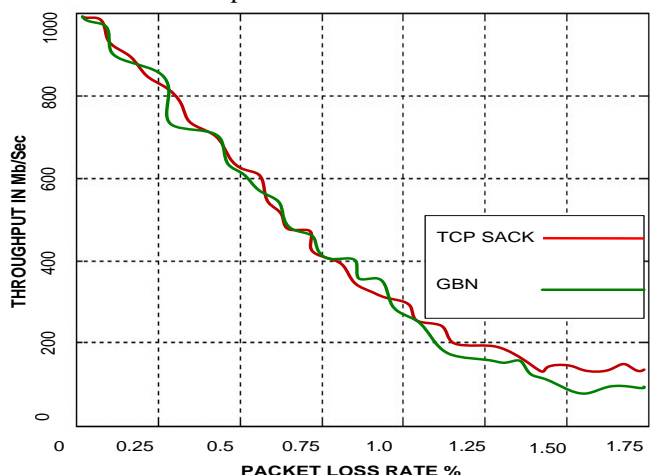
Figure 10: Packet loss vs. throughput of GBN and SACK

a. Throughput of TCP and SCTP is identical at low data rates.
b. Global business network (GBN) outperforms SACK at intermediate drop rates, for example, at 2.5% drop rate, GBN provides 25% better throughput than SACK.

However, it is different in data center and previous cases cannot be applied. We get high throughput at high drop rate.

In addition, The Bandwidth delay product does not increase considerably since at those rates with HW protocol.

The most impressive thing, which we have noticed in this context, is that GBN performs simply. Hence, it is easier to implement in HW protocol. It is clear that these experiments are helpful for evaluating these protocols in a real implementation. The findings shows real pure performance of setting and these experiments were not for revisiting practical GBN and SACK given in figure 10.

## VII. Conclusion and Future Work

In this paper, we have comprehensively studied the features of SCTP from data center point of view. We discuss the fundamental differences between the WAN and data center environments. Several issues of SCTP are discussed according to data center environment including two side implementation and protocols. In this context, we have presented research with new directions that is completely impressive. We have observed major changes on the protocol side including redesigning of streaming feature to maximize identification and provide a simple embedded acknowledgment procedure with SACK optional. We have reduced the number of M2M copies and SACK overhead and simplified the chunking data structures and TCP structure. We finally have Implemented finer grain TCP locking mechanisms for larger messages. In future, we will implement application level synchronized window flow control and utilizing topological information within a data center to improve multi-homed associations.

## Brief Biographic:

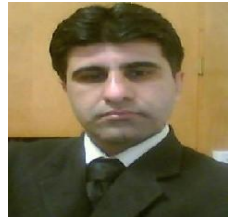

**Professor Abdul Razaque** is associated with University of Bridgeport and University of Northern Virginia, USA since 2010 as adjunct faculty Professor.

He possesses several positions at numerous international conferences and Journals including IEEE, IET, ACM, CCAIE, ICOS, ISIEA, IACSIT, Springer and Mosharka International conference. In addition he possesses fellowship form Higher Education Commission (HEC) Pakistan, University of Bridgeport, USA and Common Wealth.

Prof. Abdul Razaque served as Head of computer science department in Model colleges setup Islamabad, Pakistan from 2002 to 2009. He also led several projects as project Director for promoting the trend of information technology (IT) in Pakistan funded by United Nation organization (UNO) and World Bank during 2005 to 2008. He has authored over 80 technical papers, contributed book chapters, conducted a number of short courses and delivered invited talks, plenary lectures and presented his research more than 35 countries.

He is currently active researcher of wireless and Mobile communication (WMC) laboratory, USA and leading several interdisciplinary and collaborative projects with well-known organizations including AT&T, Nokia and Android. Prof. Abdul Razaque has chaired more than dozen of highly reputed international conferences and also delivered his lectures as Keynote Speaker. He is currently working as editor-in-chief for international Journal of Engineering and technology (IJET). His research interests include the design and development of learning environments to foster pedagogical activities, TCP/IP protocols over wireless networks, delivery of multimedia applications, ambient intelligence and wireless sensor networks.

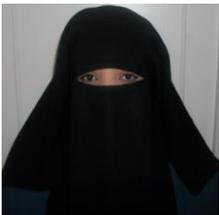

**Mrs. Fatma Almajadub** is a Master student in the School of Engineering and Computer Science at the University of Bridgeport. She has finished her Bachelor's degree at Sabha College for Computer Science in 2004-2005. She graduated with highest honors, ranking 4th in her Bachelor degree and awarded merit certificate. Mrs. Almajadub worked as a teacher at Sabha University in 2004-2009. She worked also as teacher in Algamaheria institution to high education since 2006-2009. Mrs. Almajadub is interested in programming, network area, mobile communication, and some software applications.

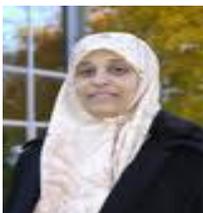

**Professor Eman Abdelfattah** received her MS and PhD Degrees in Computer Science, University of Bridgeport in May 2003 and 2011 respectively. She received "Academic Achievement Award" in Computer Science field awarded by School of Engineering, University of Bridgeport in May 2003.

Mrs. Abdelfattah worked as a programmer and computer teacher in several places. She also worked as a C++ and Java instructor in the Continuing Education Department, Housatonic Community College, Bridgeport, Connecticut. Currently, she has been working as an adjunct Professor at University of Bridgeport.

Eman has research interests in the areas of networking and communications. Her research results were published in several prestigious international conferences in networking and circuits. Eman is committee member of the various international conferences and Journals.